\documentclass[epj]{webofc}
\usepackage[utf8]{inputenc}
\usepackage[varg]{txfonts}   
\usepackage{booktabs}
\usepackage{xcolor}
\definecolor{darkred}{rgb}{0.4,0.0,0.0}
\definecolor{darkgreen}{rgb}{0.0,0.4,0.0}
\definecolor{darkblue}{rgb}{0.0,0.0,0.4}
\usepackage[bookmarks,linktocpage,colorlinks,
    linkcolor = darkred,
    urlcolor  = darkblue,
    citecolor = darkgreen]{hyperref}
\usepackage{subfigure,epstopdf,graphicx}
\usepackage{amsmath,mathrsfs,bbm,comment}
\wocname{EPJ Web of Conferences}
\woctitle{Lattice2017}
\begin{document}
%
\selectlanguage{english}
\title{%
Tensor Network study of the (1+1)-dimensional Thirring Model 
}
\author{%
\firstname{Mari Carmen} \lastname{Ba{\~n}uls}\inst{1} \and
\firstname{Krzysztof} \lastname{Cichy}\inst{2,3} \and
\firstname{Ying-Jer} \lastname{Kao}\inst{4} \and
\firstname{C.-J. David} \lastname{Lin}\inst{5} \and
\firstname{Yu-Ping} \lastname{Lin}\inst{4} \and
\firstname{David Tao-Lin} \lastname{Tan}\inst{5}\fnsep\thanks{Speaker, \email{tanlin2013.py04g@nctu.edu.tw}}
}
\institute{%
Max-Planck-Institut für Quantenoptik, Garching 85748, Germany
\and
Faculty of Physics, Adam Mickiewicz University, Poznań 61-614, Poland
\and
Institut für Theoretische Physik, Goethe-Universität Frankfurt, Frankfurt am Main 60438, Germany
\and
Department of Physics, National Taiwan University, Taipei 10617, Taiwan
\and
Institute of Physics, National Chiao Tung University, Hsinchu 30010, Taiwan
}
\abstract{%
Tensor Network methods have been established as a powerful technique for simulating low dimensional strongly-correlated systems for over two decades. Employing the formalism of Matrix Product States, we investigate the phase diagram of the massive Thirring model. We also show the possibility of studying soliton dynamics and topological phase transition via the Thirring model.
}
\maketitle
\section{Introduction}\label{intro}
Tensor Network (TN) methods \cite{ORUS2014117,Verstraete2008adv,SCHOLLWOCK201196} provide a non-perturbative tool for studying (lattice) quantum many-body systems. A common ingredient of TN algorithms is an entanglement-based ansatz for the quantum many-body wave function, which allows an efficient description of the system in spite of the exponential growth of the Hilbert space dimension with the number of lattice sites. Tensor network methods have been extremely successful in one-dimensional problems, and are becoming competitive for higher dimensional ones. They are free of the sign problem, and can be used to approximate ground states, low-lying excitations, thermal states, and (to some extent) real-time evolution. In the last years they have been applied to the Hamiltonian formulation of lattice gauge theories \cite{Tagliacozzo:2012vg,Banuls:2013jaa,Buyens:2014pga,Saito:2014bda,Silvi:2014pta,Tagliacozzo:2014bta,Banuls:2015sta,Kuhn:2015zqa,Pichler:2015yqa,Zohar:2015eda,Buyens:2015dkc,Buyens:2015tea,Saito:2015ryj,Banuls:2016lkq,Buyens:2016ecr,Banuls:2016jws,Buyens:2016hhu,Banuls:2016hhv,Silvi:2016cas,Banuls:2016gid,Zohar:2016wcf,Banuls:2017ena,Zapp:2017fcr}. 

In this article, we consider the zero-charge sector of the massive Thirring model, where the model is dual to the sine-Gordon theory \cite{PhysRevD.11.2088}. In fact, early explorations of the Thirring model suggested that the massless case is equivalent to a free scalar theory in two dimensions \cite{Johnson1961,Hagen1967,Klaiber_1969}. Without the presence of mass, a fermion-antifermion pair created from the vacuum does not separate and forms a bound state in two dimensions. Although, in general, this would not be possible if the fermions have mass, Coleman illustrated that we can still have a bosonic description for the zero-charge sector of the massive Thirring model \cite{PhysRevD.11.2088}. Since in Euclidean space the sine-Gordon theory contains topological structures, this connection allows one to simulate soliton dynamics through investigating the massive Thirring model.  

On the other hand, the sine-Gordon theory in Euclidean space is also dual to the vortex sector of the classical XY model. This model undergoes a phase transition without the appearance of long-range order in two dimensions, which is known as the Berezinskii-Kosterlitz-Thouless (BKT) transition \cite{0022-3719-6-7-010,1971JETP...32..493B}. It has been found that the topological excitations, vortices and anti-vortices, play an important role in this system. At low temperatures, a vortex forms a bound state with an anti-vortex. The vortex-anti-vortex pairs are condensed in the quasi-long-range order phase, where the correlation functions decay with the distance as a power law. At high temperatures, the system is in a disordered phase and the correlators exhibit the behavior of exponential decay. The BKT phase transition happens without breaking the $O(2)$-symmetry of the XY model. Instead, these two phases are distinguished by different scaling laws of the correlators.

This article is organized as follows. We begin by presenting the definition of the models and their relationships in Sec. \ref{sec-2}. We explain our implementation for lattice simulations in Sec. \ref{sec-3}. In Sec. \ref{sec-4}, we formulate the Matrix Product State (MPS) ansatz and the algorithm we adopt. Finally, we present our preliminary numerical results in Sec. \ref{sec-5}, and conclude in Sec. \ref{sec-6}. 

\section{Models and Dualities} \label{sec-2}
The massive Thirring model is a theory of a single Dirac field on a (1+1)-dimensional space-time, with the action defined by
\begin{equation} \label{action-thirring}
    S_{\mathrm{Th}}[\psi,\bar{\psi}] \
    = \int d^2x \left[ \bar{\psi}i \gamma^{\mu}\partial_{\mu}\psi \
      - m_{0}\,\bar{\psi}\psi \
      -\frac{g}{2} (\bar{\psi}\gamma_{\mu}\psi)^2 \right] \,,
\end{equation}
where $m_{0}$ denotes the bare mass, and $g$ is a dimensionless coupling constant. 

In \cite{PhysRevD.11.2088}, the equivalence between the zero-charge sector of the massive Thirring model and the sine-Gordon model was established. The sine-Gordon model is a theory of a single scalar field on a (1+1)-dimensional space-time, with the action defined by
\begin{equation} \label{action-SG}
    S_{\mathrm{SG}}[\phi] = \frac{1}{t}\int d^{2}x \left[ \frac{1}{2} (\partial_{\mu}\phi(x))^2 \
    + \alpha_0 \, \cos\phi(x) \right] \,,
\end{equation}
where $\alpha_0$ is a dimension-two bare coupling, and $t$ is dimensionless. The correspondence between operators and couplings is summarized by
\begin{equation} \label{bosonization}
    \bar{\psi}\gamma_{\mu}\psi \leftrightarrow \frac{1}{2\pi}\epsilon_{\mu\nu}\partial_{\nu}\phi \,,\
    \bar{\psi}\psi \leftrightarrow \frac{\Lambda}{\pi}\mathrm{cos}\phi \,,\
    \frac{4\pi}{t} = 1 + \frac{g}{\pi} \,,
\end{equation}
where $\Lambda$ is the ultra-violet cutoff. In particular, the first formula of Eq. (\ref{bosonization}) indicates the fermion number, $\int_{-\infty}^{\infty} dx_{1} \, \bar{\psi}\gamma_{0}\psi$, is equivalent to the topological charge of the sine-Gordon model, 
$
    Q = \int_{-\infty}^{\infty} dx_{1} \, \frac{1}{2\pi}\epsilon_{0\nu}\partial_{\nu}\phi \,.
$
The fermion-antifermion pairs, bound or not, are reinterpreted as the pairs of the quasi-particles of the sine-Gordon theory, namely, solitons and anti-solitons. 

It is well known that the sine-Gordon model is also dual to the vortex sector of the classical XY model described by the Hamiltonian
\begin{equation} \label{hamiltonian-XY}
    H_{\mathrm{XY}} \
    = -K \sum_{\langle ij \rangle} \cos(\theta_i - \theta_j) \,,   
\end{equation}
where $\theta$ is an angle defined on the two-dimensional lattice, and $K$ is a coupling constant. Below the critical temperature, $T_{c}$ ($T<T_{c}$), all angles are almost aligned. Yet, the long-range order is destroyed by thermal fluctuations. Despite this, the system still has a quasi-long-range order phase, with the correlation function decaying algebraically. At high temperatures ($T>T_{c}$), the system is in a disordered phase and the correlator decays exponentially. To summarize,
\begin{equation} \label{bkt-corr}
    \langle \cos(\theta_i - \theta_j) \rangle \ 
    \propto \ 
    \begin{cases}
        |i-j|^{-T/2\pi K} &\text{when $T<T_c$\,,}\\
        \exp(-|i-j|/\xi) &\text{when $T>T_c$\,,}
    \end{cases}
\end{equation}
where $\xi$ is the correlation length. In addition, we can describe the vortex sector of the XY model, which is equivalent to the Coulomb gas, by the grand canonical partition function \cite{mudry2014lecture},
\begin{equation} \label{action-XY}
     Z_{\mathrm{CG}} \left(\beta_{T} K, \beta_{T} \epsilon_{c}\right) \
        = \prod_{\mathbf{r}} \sum_{n_{\mathbf{r}}=-\infty}^{\infty} \, \ 
          exp\left( -\beta_{T} \frac{(2 \pi)^{2}K}{2} \sum_{\mathbf{r}}\sum_{\mathbf{r}'} n_{\mathbf{r}}n_{\mathbf{r}'} G_{L}(\mathbf{r}-\mathbf{r}') \ 
          - \beta_{T}\sum_{\mathbf{r}} \epsilon_{c}n_{\mathbf{r}}^{2} \right) \,,
\end{equation}
where $\beta_{T}=\frac{1}{T}$ and $G_{L}$ is the Green's function for the two-dimensional Laplacian operator. We also include the vortex density $n_{\mathbf{r}}$ and the core energy of each vortex $\epsilon_{c}$. One can find the sine-Gordon representation of Eq. (\ref{action-XY}) as,
\begin{equation}
    Z_{\mathrm{SG}} \left(\beta_{t}, \beta_{t}\alpha_{0}\right) \
        = \int \mathscr{D}\phi \, \
          exp\left( -\beta_{t} \int d^{2}x \left[ \frac{1}{2} (\partial_{\mu}\phi(x))^2 \
          - \alpha_{0} \, \cos\phi(x) \right] \right) \,,
\end{equation}
where $\beta_{t}=\frac{1}{t}$, and the sine Gordon action Eq. (\ref{action-SG}) is now written in Euclidean space (and thus the sign in front of $\alpha_0$ is flipped). The correspondence of parameters is identified by 
\begin{equation} \label{Z_SG}
\begin{aligned}
    Z_{\mathrm{SG}} \left( \frac{1}{(2\pi)^2 \beta_{T} K}, 2 e^{-\beta_{T}\epsilon_c} \right) \
    &= A \, Z_{\mathrm{CG}} \left(\beta_{T} K, \beta_{T} \epsilon_{c}\right) \,,
\end{aligned}
\end{equation}
where $A$ is a constant, and $e^{-\beta_{T}\epsilon_c}$ is known as the fugacity. Finally, we summarize the correspondence of the parameters of these three models in Table \ref{tab-1}.

\begin{table}[h!] 
    \centering
    \scalebox{1.04}{
    \begin{tabular}{| c | c | c | c |} 
        \hline
        Thirring & sine-Gordon & XY \\ \hline
        \(\displaystyle g \) & \(\displaystyle \frac{4\pi^2}{t} - \pi\) & \(\displaystyle \frac{T}{K} - \pi\) \\ \hline
        \(\displaystyle \frac{m_{0}\Lambda}{\pi} \) & \(\displaystyle \frac{\alpha_{0}}{t} \) & \(\displaystyle 2 e^{-\beta_T\epsilon_c} \) \\ \hline 
    \end{tabular}
    }
\caption{Correspondence of the parameters of the massive Thirring model, sine-Gordon theory and the classical XY model}
\label{tab-1}
\end{table}


\section{Preliminaries of the Lattice Calculation} \label{sec-3}
In our approach, the system is described by the quantum many-body wave function. To start with, we employ staggered fermions as proposed by Kogut and Susskind \cite{PhysRevD.13.1043}. In this formulation the lattice Hamiltonian derived from the action of Eq. (\ref{action-thirring}) reads 
\begin{equation}
    H_{\mathrm{Th}}^{(\mathrm{latt.})} \
        = -\frac{i}{2a} \sum_{n=0}^{N-2} 
            \Big( c^{\dagger}_{n}c_{n+1} - c^{\dagger}_{n+1}c_{n} \Big)
            + m_{0}\sum_{n=0}^{N-1} \left(-1\right)^{n} c^{\dagger}_{n}c_{n} \
        + \frac{2g}{a} \sum_{n=0}^{\frac{N}{2}-1} \
            c^{\dagger}_{2n}c_{2n} c^{\dagger}_{2n+1}c_{2n+1} \,,
\end{equation}
where $c^{\dagger}_{n}$ and $c_{n}$ are fermionic creation and annihilation operators obeying the anti-commutation relations, $\{ c^{\dagger}_{n} , c_{m} \} = \delta_{nm}$, $\{ c_{n} , c_{m} \} = 0$, $\{ c^{\dagger}_{n} , c^{\dagger}_{m} \} = 0$, and $a$ is the lattice spacing. Furthermore, the fermion variables can be mapped onto spins by using the Jordan-Wigner transformation,
\begin{equation}
\begin{aligned}
    c_{n} = exp\left( \pi i \sum_{j=1}^{n-1} S_{j}^{z} \right) S_{n}^{-} \,\,,\,\,\, c_{n}^{\dagger} = S_{n}^{+} \, exp\left( -\pi i \sum_{j=1}^{n-1} S_{j}^{z} \right) \,,
\end{aligned}
\end{equation}
with $S^{+}_{n}$ and $S^{-}_{n}$ being the ladder operators of spin-1/2. We note that this step is not strictly required, since TN can be applied to fermionic degrees of freedom. However, it is convenient to construct the full Hamiltonian using the Pauli matrices. Therefore, the spin Hamiltonian we use in the simulation is given by
\begin{equation} \label{hamiltonian-spin}
\begin{aligned} 
    H_{\mathrm{spin}} \
        =& -\frac{1}{2a} \sum_{n=0}^{N-2} \left( S_{n}^{+}S_{n+1}^{-} 
            + S_{n+1}^{+}S_{n}^{-} \right)
            + m_{0} \sum_{n=0}^{N-1} \left(-1\right)^{n} \left( S_{n}^{z}+\frac{1}{2} \right) &\\
        &+ \frac{2g}{a} \sum_{n=0}^{\frac{N}{2}-1} \left( S_{2n}^{z}+\frac{1}{2} \right) \ 
            \left( S_{2n+1}^{z}+\frac{1}{2} \right) \,.
\end{aligned}
\end{equation}
We will explain how to solve the Hamiltonian numerically in Sec. \ref{sec-5}. In addition, the total $S^z$ polarization $S_{\mathrm{tot}}^z=\sum_n S_n^z$ is a conserved quantity, related to the charge in the fermionic language. We can thus explore the zero-charge sector by requiring $S_{\mathrm{tot}}^z=0$. Further details will be illustrated in the next section. 


\section{Tensor Network Methods} \label{sec-4}
The concept of entanglement is central to the Tensor Network states and the algorithms based on them. Any pure state $|\Psi\rangle$ of a composite Hilbert space $\mathscr{H} = \mathscr{H}_{A} \otimes \mathscr{H}_{B}$, $\mathscr{H}_{A} = \mathbb{C}^{d_{A}}$ and $\mathscr{H}_{B} = \mathbb{C}^{d_{B}}$, has a Schmidt decomposition,
\begin{equation}
    |\Psi\rangle = \sum_{i=1}^{r} s_{i}\,|i_{A}\rangle |i_{B}\rangle \,,
\end{equation}
where $s_{i}$'s, known as Schmidt coefficients, are positive, monotonically decreasing real numbers which determine the entanglement with respect to the bipartition $A|B$, and $r \leq min(d_{A},d_{B})$. We can compute the entropy of entanglement 
\begin{equation} \label{bipartite_entanglement_entropy}
    S_{A|B} = -Tr\,(\rho_{A}\log\rho_{A}) = -\sum_{i=1}^{r} s_{i}^{2} \log s_{i}^{2} \,,
\end{equation}
where $\rho_A$ is the reduced density matrix for the subsystem $A$. If the system is weakly entangled, the Schmidt coefficients often decrease very fast. This provides us with a way to approximate the state by introducing an entanglement cutoff which truncates the number of the Schmidt coefficients from $r$ to $r'<r$. 

\begin{figure}[tp]
   \centering
   \subfigure[The MPS.]%
             {\includegraphics[width=0.26\textwidth,clip]{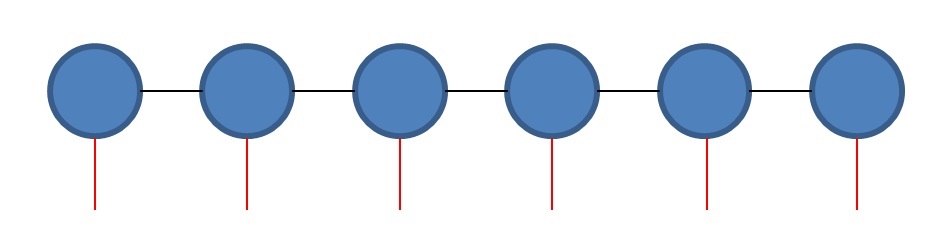}\label{fig-mps}}
   \subfigure[The MPO.]%
             {\includegraphics[width=0.26\textwidth,clip]{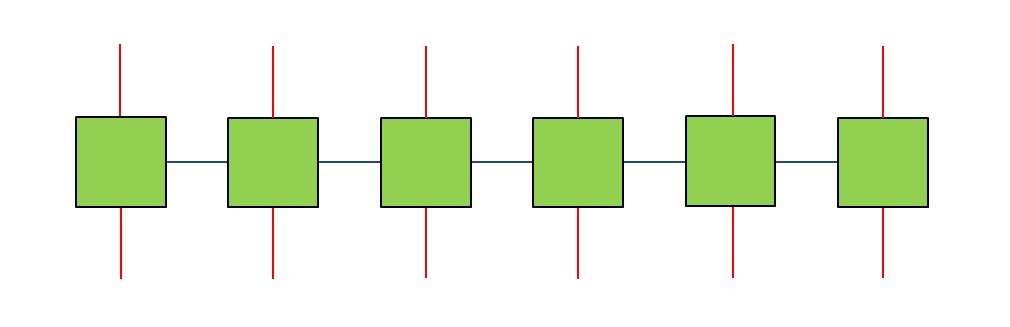}\label{fig-mpo}}
   \subfigure[The expectation value of the MPO (b) in the state represented by the MPS (a).]%
             {\includegraphics[width=0.26\textwidth,clip]{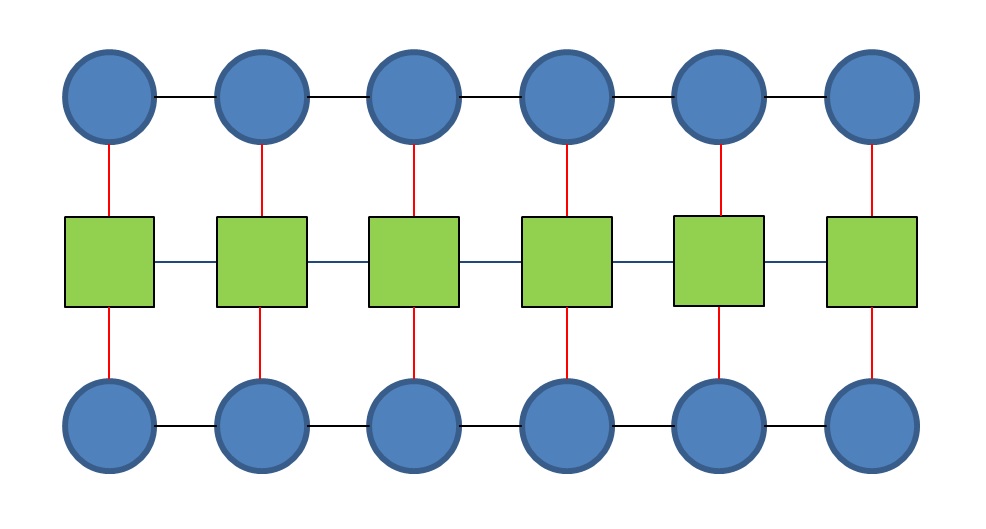}\label{fig-dmrg}}
   \caption{The graphical representation of Tensor Network states and algorithms based on it.}
\end{figure}

\subsection{Matrix Product States}
For a $N$-partite system, with Hilbert space $\mathscr{H} = \mathscr{H}_{0} \otimes \mathscr{H}_{1} \otimes \cdots \otimes \mathscr{H}_{N-1}$, $\mathscr{H}_{n} = \mathbb{C}^{d}$ for every $n$, a MPS is a state of the form \cite{SCHOLLWOCK201196}
\begin{equation}
    |\Psi\rangle = \sum_{a_{0},a_{1},\cdots} M[0]_{a_{0}}^{\sigma_{0}} M[1]_{a_{0},a_{1}}^{\sigma_{1}} \cdots M[N-1]_{a_{N-2}}^{\sigma_{N-1}}\,|\sigma_{0}\sigma_{1}\cdots\sigma_{N-1}\rangle \,,
\end{equation}
where $M[n]^{\sigma_n}$ are $D\times D$ matrices for $0<n<N-1$, and vectors of dimension $D$ for $n=0,\, N-1$ (see Fig. \ref{fig-mps}). The parameter $D$ is called the bond dimension. Any state of the Hilbert space can be written as a MPS for $D\leq d^{\lfloor N/2 \rfloor}$, while restricting $D$ to a small value plays the role of a cutoff in the Schmidt rank.

\subsection{Density Matrix Renormalization Group}
Our objective is to find a MPS approximation to the ground state (and excitations) of the model. This can be achieved by a variational minimization of the energy over the set of MPS with fixed bond dimension $D$, similar to the Density Matrix Renormalization Group (DMRG) algorithm \cite{SCHOLLWOCK201196}. The algorithm is simplified when the Hamiltonian is written as a Matrix Product Operator (MPO), i.e. a MPS in the operator vector space \cite{ORUS2014117,Verstraete2008adv}, see Fig. \ref{fig-mpo}.
\begin{equation}
    H = \sum_{b_{0},b_{1},\cdots,b_{N-2}} W[0]_{b_{0}}^{\sigma_{0},\sigma'_{0}}W[1]_{b_{0},b_{1}}^{\sigma_{1},\sigma'_{1}} \cdots W[N-1]_{b_{N-2}}^{\sigma_{N-1},\sigma'_{N-1}} |\sigma_{0}\sigma_{1}\cdots\sigma_{N-1}\rangle\langle\sigma'_{0}\sigma'_{1}\cdots\sigma'_{N-1}| \,,
\end{equation}
where $W[n]$ are $\chi\times \chi$ matrices ($\chi$ dimensional vectors for the edges, since we adopt open boundary conditions).
We start with expressing the functional 
\begin{equation} \label{variational principle}
    E\left[\Psi\right] = \frac{\langle\Psi|H|\Psi\rangle}{\langle\Psi|\Psi\rangle}
\end{equation}
in terms of a Tensor Network as Fig. \ref{fig-dmrg}. If the $M[n]$ tensors for all but one site, $k$, are fixed, Eq. (\ref{variational principle}) can be optimized over $M[k]$ by solving a generalized eigenvalue equation \cite{SCHOLLWOCK201196}. This can be repeated for all sites, from left to right and back, until sufficient convergence of $E/N$ is achieved.

The Hamiltonian in Eq. (\ref{hamiltonian-spin}) conserves the total $S^z$ polarization $S_{\mathrm{tot}}^z=\sum_n S_n^z$. It is thus convenient to restrict the search to a sector of specific $S_{\mathrm{tot}}^z=S_{\mathrm{target}}$. To this end we add a penalty term \cite{Banuls:2013jaa} to the Hamiltonian
\begin{equation} \label{hamiltonian-spin_penalty}
    H_{\mathrm{spin}}^{(\mathrm{penalty})} = H_{\mathrm{spin}} + \lambda \left( \sum_{n=0}^{N-1} S_{n}^{z} - S_{\mathrm{target}} \right)^2 \,,
\end{equation}
where the magnitude of $\lambda$ should be chosen to be large enough to ensure that all the states with $\langle\Psi|\sum_{n=0}^{N-1} S_{n}^{z}|\Psi\rangle \neq S_{target}$ have energy above the lowest state in the desired sector. The MPO representation of the Hamiltonian of Eq. (\ref{hamiltonian-spin_penalty}) is therefore given by
\begin{equation} \label{mpo-edge}
\begin{aligned}
    & W[0]= 
    \begin{pmatrix}
        \mathbbm{1} & -\frac{1}{2a} S^{+} & -\frac{1}{2a} S^{-} & 2\lambda S^{z} & \frac{2g}{a} S^{z} \, 
        & \beta_{n}S^{z}+\gamma\mathbbm{1} 
    \end{pmatrix} \,, &\\
    & W[N-1]= 
    \begin{pmatrix}
    \beta_{n}S^{z}+\gamma\mathbbm{1} & S^{-} & S^{+} & S^{z} & S^{z} & \mathbbm{1}
    \end{pmatrix}^T \,,
\end{aligned}
\end{equation}
for the tensors at the boundaries and
\begin{equation} \label{mpo}
    W[n]= 
    \begin{pmatrix}
        \mathbbm{1} & -\frac{1}{2a} S^{+} & -\frac{1}{2a} S^{-} & 2\lambda S^{z} & 
        \frac{2g}{a}\frac{1+(-1)^n}{2}S^z &
        \beta_{n}S^{z}+\gamma\mathbbm{1} \\
        0 & 0 & 0 & 0 & 0 & S^{-} \\
        0 & 0 & 0 & 0 & 0 & S^{+} \\
        0 & 0 & 0 & \mathbbm{1} & 0 & S^{z} \\
        0 & 0 & 0 & 0 & 0 & S^{z} \\
        0 & 0 & 0 & 0 & 0 & \mathbbm{1}
    \end{pmatrix} \,,
\end{equation}
\\
for the remaining sites, where
\begin{equation}
    \beta_{n} = \frac{g}{a} + \left(-1\right)^{n} m_{0} - 2\lambda \, S_{\mathrm{target}} \,\,,\,\,\, \
    \gamma = \lambda \left( \frac{1}{4} + \frac{S_{\mathrm{target}}^{2}}{N} \right) + \frac{g}{4a} \,.
\end{equation}


\section{Results} \label{sec-5}
\begin{figure}[bp]
   \centering
   \subfigure[]%
             {\includegraphics[width=0.46\textwidth,clip]{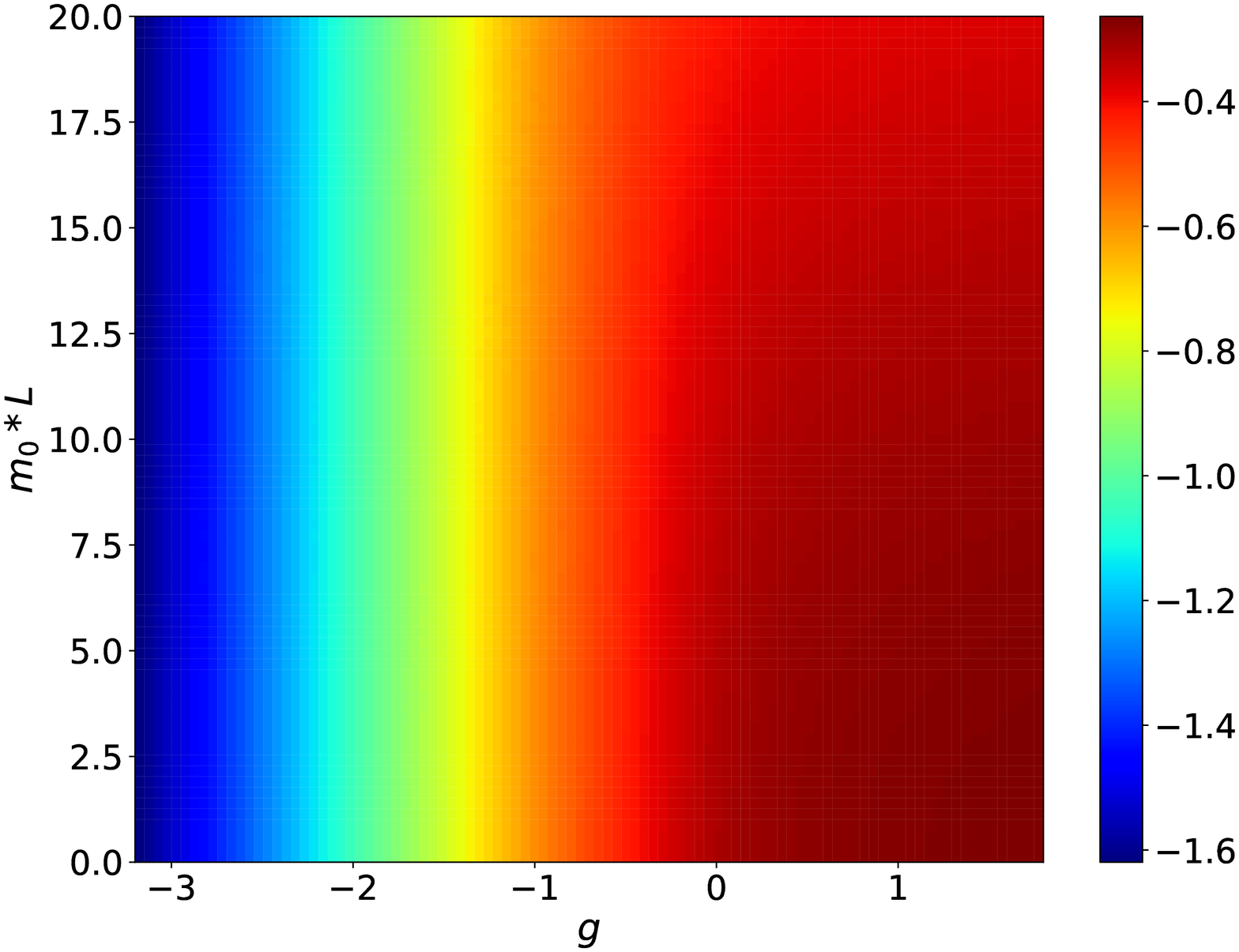}\label{fig-N40-E-sz0}}
   \subfigure[]%
             {\includegraphics[width=0.46\textwidth,clip]{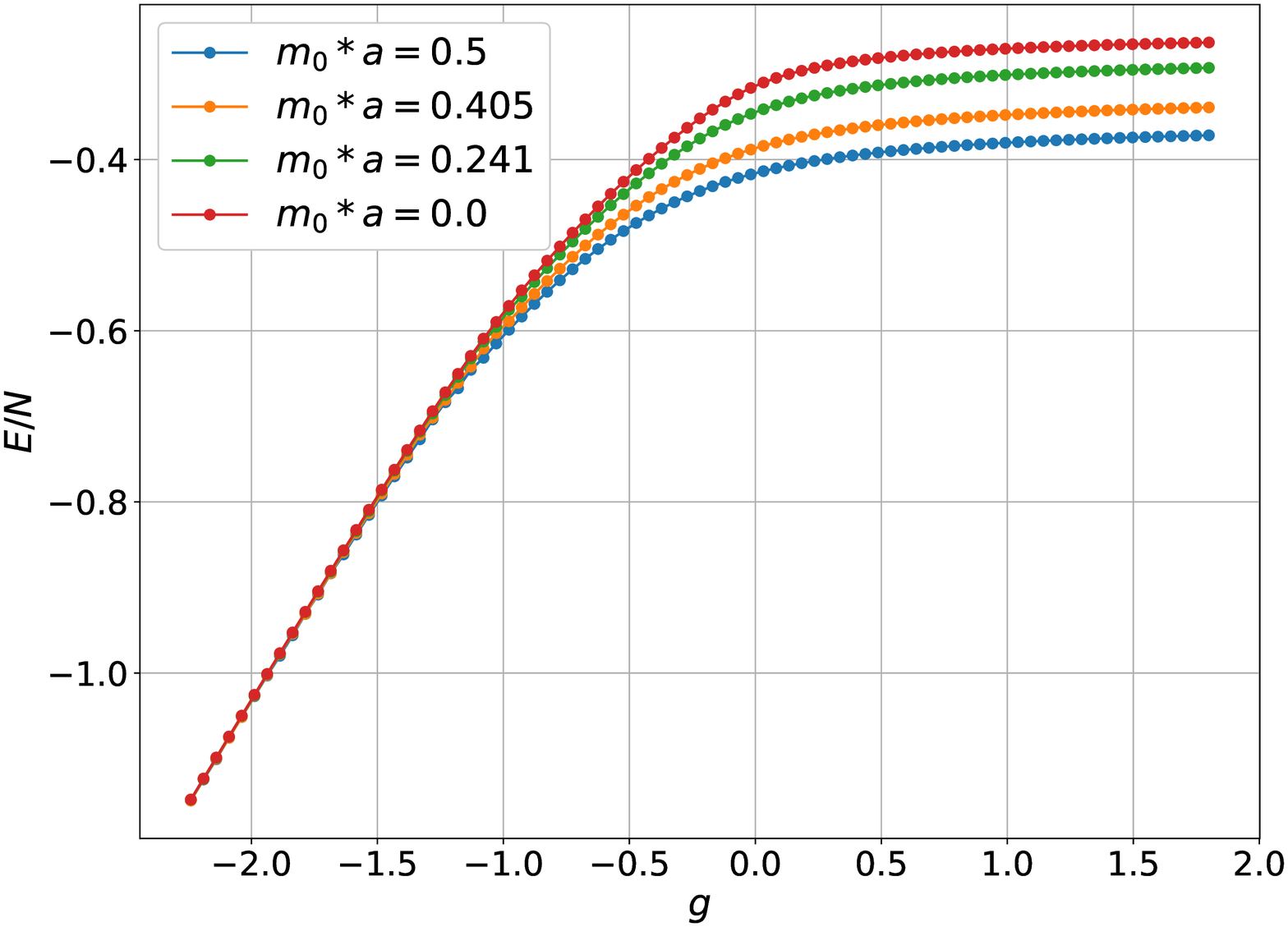}\label{fig-N40-E-mas}}
   \caption{(a) GS energy density $E/N$ of the Hamiltonian Eq. (\ref{hamiltonian-spin_penalty}), with $S_{\mathrm{target}}=0$. (b) GS energy density $E/N$ for different $m_0~a$'s. The system size is 40 sites, with bond dimension $D=100$ for both (a) and (b).} 
   \label{fig-energy}
\end{figure}

In this section, we present our results obtained in the presence of the penalty term with $S_{\mathrm{target}}=0$. Figure~\ref{fig-energy} illustrates our investigation of the ground state (GS) energy, extracted using the DMRG algorithm as described in Sec.~\ref{sec-4}, for system size $N=40$ and bond dimension $D=100$. The plot in Fig.~\ref{fig-N40-E-sz0} shows a scan on the $(g, m_{0}L)$ plane, with $L$ denoting the spatial volume of the lattice. In the regime $g < 0$, it is evident that the GS energy decreases when $g$ turns more negative. This is in accordance with Coleman's argument that the GS energy is always negative and scales as $\lambda^{1/(1+g/\pi)}$, where $\lambda$ is a dilatation parameter \cite{PhysRevD.11.2088}. Figure~\ref{fig-N40-E-mas} exhibits the GS energy per site ($E/N$) as a function of $g$.  We observe that when $g \lesssim -1.5$,  $E/N$ is almost independent of the mass parameter $m_{0}$.  To understand this behavior, we notice that the renormalization group (RG) analysis of the sine-Gordon theory concludes that the ($\mathrm{cos}\phi$) operator is irrelevant when $t > 8\pi$. This corresponds to the region $g < -\pi /2$. In other words, the Thirring model is indeed describing a free bosonic theory in this regime. This can lead to the observed independence of $m_0$.

\begin{figure}[tp]
   \centering
   \subfigure[]%
             {\includegraphics[width=0.46\textwidth,clip]{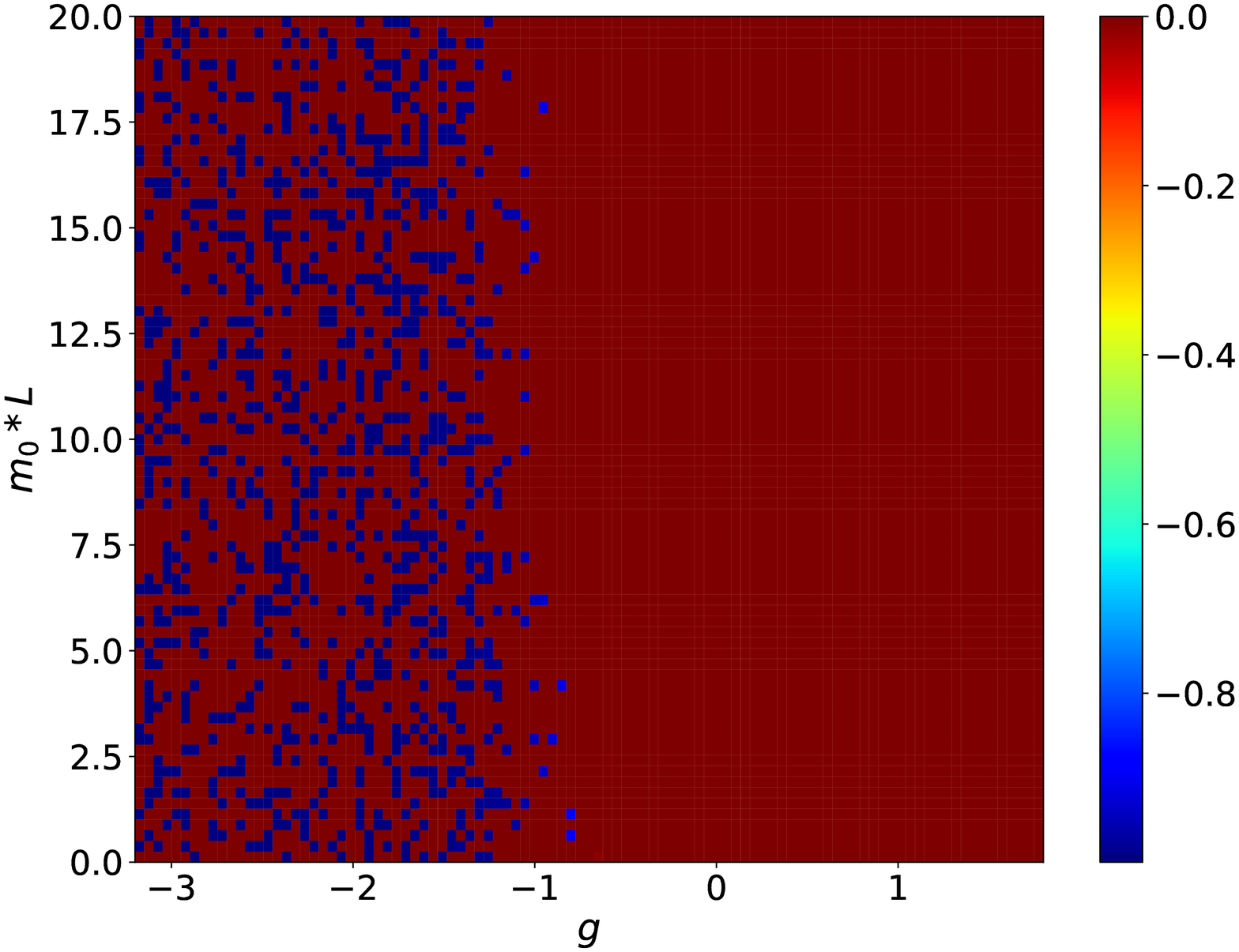}\label{fig-N40-sector-sz0}}
   \subfigure[]%
             {\includegraphics[width=0.46\textwidth,clip]{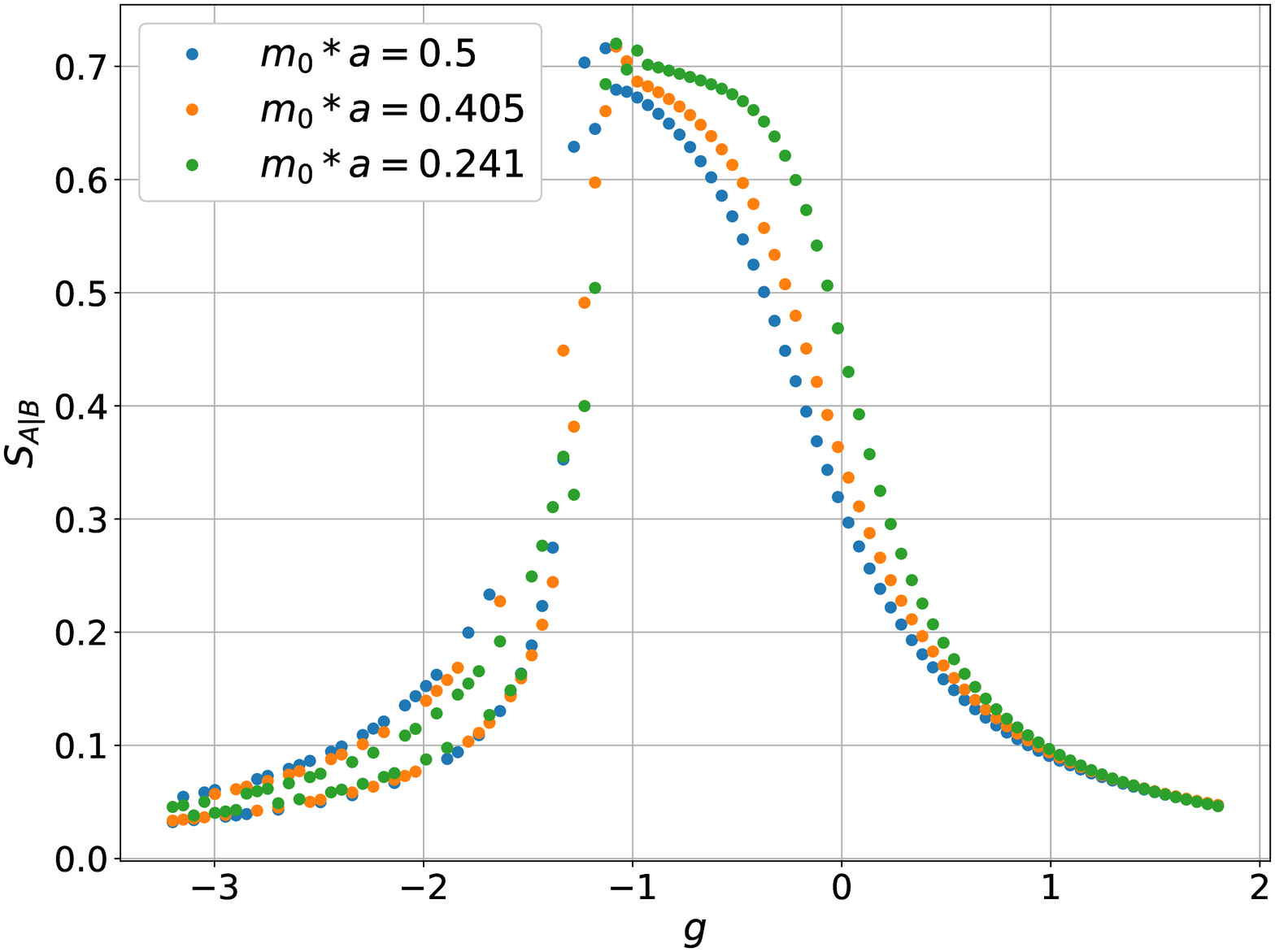}\label{fig-N40-entropy}}
   \caption{(a) $S^z_{\mathrm{tot}}$ sector of the Hamiltonian Eq. (\ref{hamiltonian-spin_penalty}), with $S_{\mathrm{target}}=0$. (b) The bipartite entanglement entropy, with the system divided into two equal-size subsystems, $A$ and $B$. The system size is 40 sites, with bond dimension $D=100$ for both (a) and (b).}
   \label{fig-Sz_and_entropy}
\end{figure}

To further examine the convergence of the algorithm, we also check $S_{\mathrm{tot}}^z$ and the bipartite entanglement entropy computed using the GS extracted from our DMRG simulations. To calculate the bipartite entanglement entropy, we divide the system into two equal-size subsystems, $A$ and $B$, and then determine the entanglement entropy $S_{A|B}$ using Eq.~(\ref{bipartite_entanglement_entropy}). Figure~\ref{fig-Sz_and_entropy} displays results from this study. We notice that in Fig.~\ref{fig-N40-sector-sz0} the GS does not always
converge to the desired $S_{\mathrm{tot}}^{z}$ sector, even in the presence of the penalty term with $S_{\mathrm{target}} = 0$. This happens when $g \lesssim -1$. In addition, it is observed in Fig.~\ref{fig-N40-entropy} that $S_{A|B}$ shows unstable, scattering behavior. In order to understand this instability, we perform DMRG simulations starting from a GS with $S_{\mathrm{tot}}^{z} = 0$ at $g < -1$, and slowly changing $g$. Namely, we use the ground state with $S_{\mathrm{tot}}^{z} = 0$ obtained at a value of $g$ as our initial MPS in the DMRG computation for another choice of $g$. With this procedure, we find that the instability demonstrated in Fig.~\ref{fig-Sz_and_entropy} disappears. Presently we are carrying out further investigation of this issue.

\section{Conclusion and Outlook} \label{sec-6}
In this project, we implemented the MPS method for the massive Thirring model on the lattice, employing staggered fermions. This article presents our exploratory numerical results from this work.  Our initial findings show that the DMRG computation for this model can be performed.  Using the dualities described in Sec.~\ref{sec-2}, we will investigate the topological phase transition and the soliton dynamics in the XY model and in the sine-Gordon theory.

It is well known that there is a BKT-type phase transition at $T \sim \pi K /2$ in the XY model. From Table~\ref{tab-1}, it means that this transition can be observed at $g \sim -\pi /2$ in the massive Thirring model. The issue is to use the dualities and the Jordan-Wigner transformation to write the XY-model vortex-anti-vortex correlators in the format of MPO and MPS. This aspect of the project is now under investigation, and will appear in a separate publication in the near future \cite{in-preparation}.


\section*{Acknowledgments}
KC was supported in part by the Deutsche Forschungsgemeinschaft (DFG), project nr. CI 236/1-1. CJDL and DTLT acknowledge the support from Taiwanese MoST via grant 105-2628-M-009-003-MY4, and are grateful for the hospitality of the University of Frankfurt during the progress of this work. YJK and YPL acknowledge the support from Taiwanese MoST via grant 105-2112-M-002 -023 -MY3, 104-2112-M-002 -022 -MY3.

\bibliography{lattice2017}

\end{document}